\documentclass[aps]{revtex4-2}\usepackage{graphicx}
\usepackage{bm}
\usepackage{amsmath}

\begin{document}

\title{Spontaneous symmetry and antisymmetry breaking in a ring with two
potential barriers}
\author{Hidetsugu Sakaguchi$^{1}$, Boris A. Malomed$^{2,3}$\footnote{The corresponding author; e-mail malomed\@tauex.tau.ac.il}, and T. J. Taiwo$%
^{4}$}
\address{$^{1}$Department of Applied Science for Electronics and Materials, Interdisciplinary Graduate School of
Engineering Sciences, Kyushu University, Kasuga, Fukuoka 816-8580, Japan}
\address{$^{2}$Department of Physical Electronics, School of Electrical Engineering,
Faculty of Engineering, and Center for Light-Matter Interaction, Tel Aviv
University, P.O. Box 39040 Tel Aviv, Israel}
\address{$^{3}$Instituto de Alta Investigaci\'{o}n, Universidad de Tarapac\'{a}, Casilla 7D,
Arica, Chile}
\address{$^{4}$Department of Physics, United Arab Emirates
University. Al Ain, United Arab Emirates}

\begin{abstract}
We propose a fundamental setup for the realization of spontaneous symmetry
breaking (SSB) and spontaneous antisymmetry breaking (SASB) in the framework
of the nonlinear Schr\"{o}dinger equation with the self-attractive and
repulsive cubic term, respectively, on a one-dimensional ring split in two
mutually symmetric boxes by delta-functional potential barriers, placed at
opposite points. The system is relevant to optics and BEC. The spectrum of
the linearized system is found in analytical and numerical forms. SSB and
SASB are predicted by dint of the variational approximation, and studied in
the numerical form. A particular stable solution, which demonstrates strong
asymmetry, is found in an exact form. In the system with the attractive
nonlinearity, the SSB of the symmetric ground state is initiated by the
modulational instability. It creates stationary asymmetric states through a
supercritical bifurcation. In the self-repulsive system, SASB makes the
lowest antisymmetric excited state unstable, transforming it into an
antisymmetry-breaking oscillatory mode.
\end{abstract}

\maketitle

\textbf{Spontaneous symmetry breaking (SSB) is a fundamental aspect of
complexity in\ a broad range of nonlinear dynamical systems. While
eigenstates of linear systems with intrinsic symmetry (which may be
typically represented by a double-well potential, or by a symmetric pair of
subsystems coupled by linear mixing) exactly follow this intrinsic
structure, featuring either symmetry or antisymmetry with respect to it (in
particular, this is a fundamental feature of quantum mechanics, where the
ground state (GS) and all excited states (ESs) of even orders precisely
follow the symmetry of the underlying potential, while the first ES and all
others of odd orders are antisymmetric), a\ sufficiently strong
self-attractive nonlinearity may fundamentally change the situation by
destabilizing the symmetric GS and replacing it by a stable \textit{%
asymmetric} one. The self-repulsive nonlinearity does not break the symmetry
of the GS, but it may destabilize the lowest antisymmetric ES. The latter
effect may be called spontaneous antisymmetry breaking (SASB). The SSB/SASB
phenomenology was studied in detail theoretically and demonstrated
experimentally in many settings, especially in nonlinear optics and
photonics and in quantum matter (primarily, ultracold atomic gases in the
state of the Bose-Einstein condensation (BEC)). Nevertheless, it remains a
relevant objective to explore SSB and SASB in the most basic settings, where
the corresponding phase transitions (alias bifurcations) may be demonstrated
in a transparent form. This work introduces a basic setting which was not
studied before, \textit{viz}., a ring split in two mutually symmetric boxes
by a pair of identical narrow potential barriers set at diametrically
opposite points and represented by delta-functions. The model is represented
by the nonlinear Schr\"{o}dinger equation with the pair of delta-functional
repulsive potentials and self-attractive or repulsive cubic nonlinearity. In
terms of the physical realization, the model is relevant to optics and BEC
alike. The relative simplicity of the model makes it possible to develop a
systematic analytical consideration, by means of the variational
approximation, which is performed parallel to a comprehensive numerical
analysis. In addition to that, a particular exact analytical solution is
found. Both the analytical and numerical results reveal the \textit{%
supercritical bifurcation} representing the SSB\ in the case of the
attractive nonlinearity. The bifurcation destabilizes the symmetric GS,
replacing it by an explicitly found asymmetric one. In the case of the
self-repulsion, the SASB destabilizes the lowest antisymmetric ES. However,
no stationary state with broken antisymmetry is found in the latter case;
instead, the basic outcome of the SASB is replacement of the lowest ES by a
breather, i.e., a robust oscillatory state with broken antisymmetry.}

\section{Introduction}

Symmetry is a profound property of many physical systems, therefore
spontaneous symmetry breaking (SSB) in nonlinear systems is a fundamental
phenomenon with well-known manifestations in optics and photonics, matter
waves (Bose-Einstein condensates, BECs), and other fields \cite{book}.
Following early theoretical works \cite{Davies,Eilbeck}, the SSB
phenomenology was studied in detail theoretically in dual-core optical
fibers (alias nonlinear couplers), especially as concerns its realization
for solitons \cite{Wabnitz}-\cite{Mostofi}, see also an early review \cite%
{Wabnitz-review} and an updated one \cite{Peng-book}. These theoretical
models are based on linearly-coupled nonlinear Schr\"{o}dinger (NLS)
equations. The experimental demonstration of SSB and inter-core switching
for solitons in dual-core fibers was reported essentially later \cite{Ignac}%
. Subsequently, this effect was predicted in photonic lattices (where it was
also realized in the experiment) \cite{Zhigang}, see also Refs. \cite{Panos}
and \cite{Savona}, metamaterials \cite{Kivshar}, and dielectric resonators
\cite{resonators}. It was also experimentally demonstrated in coupled lasers
\cite{lasers,nano,plasmon-lasers}. In the latter case, SSB happens in
dissipative nonlinear photonic systems, that may be modeled by linearly
coupled complex Ginzburg-Landau equations \cite{Sigler}.

Systems with parity-time ($\mathcal{PT}$) symmetry represent the border
between conservative and dissipative ones. Peculiarities of the spontaneous
breaking of the $\mathcal{PT}$ symmetry were analyzed theoretically \cite%
{Christo,Radik,Barash,PT breaking,RMP,Dmitriev} and demonstrated
experimentally \cite{Segev}. Most recently, SSB of solitons was analyzed for
linearly coupled systems of NLS equations with fractional diffraction, that
can be implemented in appropriately designed optical cavities \cite{Strunin}%
. In the context of BEC, SSB and related Josephson oscillations were
predicted \cite{Walls}-\cite{we2}\ and experimentally demonstrated \cite%
{Oberthaler} in condensates (as well as in other setups using cold atoms
\cite{Korea}) loaded in double-well potential traps. SSB of solitons was
also studied in systems of coupled Korteweg -- de Vries equations, which may
be realized in fluid dynamics \cite{Jorge}. Manifestations of SSB were
predicted as well in completely different physical settings, such as an
ensemble of self-propelled particles \cite{Eshel}.

The SSB driven by the self-attractive (focusing) nonlinearity leads to
destabilization of the system's spatially symmetric ground state (GS) and
its replacement by a GS featuring a stationary asymmetric spatial structure.
Under the action of the repulsive (defocusing) nonlinearity, the symmetric
GS always remains stable, but the lowest excited state (ES), which is
spatially antisymmetric in linear systems, may be destabilized if the
self-repulsion is strong enough, which leads to the effect of the
spontaneous antisymmetry breaking\textit{\ }(SASB) \cite{Michal,Raymond}.

It is relevant to stress that the SSB concept considered in the
above-mentioned works is completely different from what is also called SSB
in the field theory \cite{fields1,fields2,phi-4,fields3}, where it is the
basis of the Higgs mechanism responsible for the creation of masses of
particles \cite{Higgs1,Higgs2,Higgs3}.

The theoretical analysis of SSB is based on numerical methods, which may be
efficiently combined with the analytical approach in the form of the
variational approximation (VA) \cite%
{Pare,Maimistov,Skinner,Panos,we,Michal,Carretero}. In this connection, it
is relevant to introduce the most basic models, which make it possible to
explore the SSB\ phenomenology in a transparent form. In particular, Ref.
\cite{Thawatchai} proposed a one-dimensional (1D) NLS equation for the
complex wave field $u(x)$, with the self-attractive cubic nonlinearity
represented by a pair of delta-functions, $\left[ \delta \left( x+a/2\right)
+\delta \left( x-a/2\right) \right] |u|^{2}u$, with separation $a$ between
them. This model makes it possible to analyze the SSB in an exact analytical
form. Also natural is to consider a 1D ring with the attractive nonlinearity
concentrated at a pair of diametrically opposite points \cite{Han Pu}. It
was found that the latter model does not give rise to SSB if the spatial
structure of the nonlinearity is represented by ideal delta-functions;
nevertheless, the SSB\ takes place if the delta-functions are replaced by
Gaussian profiles of a finite width \cite{Han Pu}.

While the ring with the self-attractive nonlinearity concentrated at two
opposite points seems as a rather \textquotedblleft exotic" (although
physically possible \cite{Azbel}) configuration, in this work we aim to
introduce a more natural setting based on a ring which is split in two
mutually symmetric boxes (half-rings) by the linear\emph{\ }repulsive
potential represented by two delta-functions placed at diametrically
opposite points. With the usual uniform self-focusing or defocusing
nonlinearity, the respective NLS equation takes the form of
\begin{equation}
i\frac{\partial u}{\partial z}+\frac{1}{2}\frac{\partial ^{2}u}{\partial
x^{2}}+\sigma |u|^{2}u=\epsilon \left[ \delta \left( x+\frac{\pi }{2}\right)
+\delta \left( x-\frac{\pi }{2}\right) \right] u,  \label{NLS}
\end{equation}%
where $\epsilon >0$ is the strength\ of the potential barriers that split
the ring. This equation is written in the form adopted in optics, with
propagation distance $z$ playing the role of the evolutional variable \cite%
{Agrawal}, while the coordinate $x$ runs along the ring, taking values
\begin{equation}
-\pi \leq x\leq +\pi  \label{x}
\end{equation}%
and imposing the periodic boundary conditions
\begin{equation}
u(x=-\pi )\equiv u\left( x=+\pi \right) ,\ \frac{\partial u}{\partial x}%
(x=-\pi )\equiv \frac{\partial u}{\partial x}\left( x=+\pi \right) .
\label{bc}
\end{equation}%
Further, coefficients $\sigma =+1$ and $-1$ correspond, respectively, to the
focusing (attractive) and defocusing (repulsive) sign of the nonlinearity in
Eq. (\ref{NLS}). Stationary solutions to Eq. (\ref{NLS}) with real
propagation constant $k$ are looked for below in the usual form, $u\left(
x,t\right) =\exp \left( ikz\right) U(x)$, where real function $U(x)$
satisfies the stationary real equation,%
\begin{equation}
-kU+\frac{1}{2}\frac{d^{2}U}{dx^{2}}+\sigma U^{3}=\epsilon \left[ \delta
\left( x+\frac{\pi }{2}\right) +\delta \left( x-\frac{\pi }{2}\right) \right]
U.  \label{U}
\end{equation}

The repulsive potential in the form of the single delta-function may
introduce SSB if it is placed at the midpoint, $x=0$, of an infinitely deep
potential box of a finite size $a$, which is described by the NLS equation
with zero boundary conditions at points $x=\pm a/2$ \cite{NatPhot,Shamriz}.
Note that the box potentials, induced by means of an optical technique, are
available in experiments with BEC \cite{box}.

The split-ring model, based on Eqs. (\ref{NLS})-(\ref{bc}), admits
straightforward interpretations in terms of optics and matter waves alike.
In the former case, Eq. (\ref{NLS}) governs the propagation of light along a
cylindrical surface (in particular, in hollow fibers and waveguides \cite%
{hollow0,hollow1,hollow2}). In this case, the potential barriers represent
narrow strips of a material with a lower refractive index, running along the
waveguide. In the application to matter waves, Eq. (\ref{NLS}), with $z$
replaced by time $t$, is the scaled Gross-Pitaevskii equation which governs
the evolution of the mean-field wave function $u\left( x,t\right) $ of BEC
loaded in a toroidal trap. Such setups were created in experiments,
combining magnetic, optical, and radiofrequency trapping potentials \cite%
{magnetic-torus,Phillips,averaged-torus,torus}. The narrow potential
barriers in the corresponding ring configuration can be induced by a
blue-detuned optical sheet cutting the torus, which has been realized in the
experiment as well \cite{weak-link}.

The following presentation is arranged as follows. First, the spectrum of
the linearized version of Eq. (\ref{NLS}) (with $\sigma =0$) is presented in
Section 2. In the framework of the full nonlinear model, the SSB and SASB
effects are studied, severally, for nonlinear states which originate from
the spatially even GS and lowest spatially odd ES of the linear spectrum.
Namely, SSB of the GS, caused by the action of the self-focusing ($\sigma =+1
$ in Eq. (\ref{NLS})), is addressed in Section 3, and SASB of the lowest ES
under the action of the self-defocusing ($\sigma =-1$) is considered in
Section 4. The SSB is illustrated, first, by a particular exact analytical
solution. Then, generic configurations produced by the SSB and their
dynamics (including stability) are studied, in parallel, by means of the
analytical VA and systematically employed numerical methods, which
demonstrates good accuracy of the VA (in addition, an analytical
approximation for large $\epsilon $ is developed too). In the case of the
self-focusing nonlinearity we identify the SSB bifurcation as one of the
forward (alias supercritical \cite{Iooss,Knobloch}) type. Past the
bifurcation point, the emerging state with the broken symmetry is stable,
while the symmetric one develops instability which initiates oscillatory
dynamics, that features periodically recurring dynamical symmetry breaking.
The instability combines features of SSB and modulational instability (MI),
which is well known in 1D systems with the cubic self-focusing \cite%
{Agrawal,MI-BEC,MI-optics}.

On the other hand, the variational and numerical analysis of SASB\
demonstrates that, past the point of the destabilization of the lowest ES,
no stationary states with broken antisymmetry emerge. Instead, the
simulations demonstrate establishment of a robust antisymmetry-breaking
oscillatory state The paper is concluded by Section 5.

\section{The spectrum of the linear system}

The linearized form of Eq. (\ref{U}),
\begin{equation}
-kU=-\frac{1}{2}\frac{d^{2}U}{dx^{2}}+\epsilon \left[ \delta \left( x+\frac{%
\pi }{2}\right) +\delta \left( x-\frac{\pi }{2}\right) \right] U,
\label{Ulin}
\end{equation}%
takes the form of the stationary quantum-mechanical Schr\"{o}dinger equation
with the repulsive delta-functional potential barriers and formal energy
eigenvalue $-k$. The effect of the barriers amounts to the jump of the
derivative of $U(x)$ at points $x=\pm \pi /2$, while $U(x)$ itself is
continuous at these points:%
\begin{equation}
\frac{dU}{dx}\left( x=\pm \frac{\pi }{2}+0\right) -\frac{dU}{dx}\left( x=\pm
\frac{\pi }{2}-0\right) =2\epsilon U\left( x=\pm \frac{\pi }{2}\right) .
\label{jump}
\end{equation}%
To produce numerical solutions, the delta-functions were replaced, as usual,
by narrow Gaussians, with width $\Delta x=0.01$.

Eigenstates of Eq. (\ref{Ulin}) may be defined as ones which are odd or even
with respect to the positions of the delta-functions, $x=\pm \pi /2$. In the
former cases, these are%
\begin{equation}
U_{\mathrm{odd}}(x)=U_{0}\sin \left( \sqrt{-2k}\left( x+\frac{\pi }{2}%
\right) \right) ,  \label{odd}
\end{equation}%
with arbitrary amplitude $U_{0}$. Because $U_{\mathrm{odd}}(x)$ vanishes at $%
x=\pm \pi /2$, the delta-functional potential barriers produce no effect on
these eigenstates, hence the respective spectrum of eigenvalues $k$ is the
same as in the free-space ring, determined solely by the periodicity
conditions (\ref{bc}):%
\begin{equation}
k_{\mathrm{odd}}=-\left( n+1\right) ^{2}/8,~n=1,3,5,...,  \label{k-odd}
\end{equation}%
where the odd quantum number $n$ is the ES order.

The eigenstates which are even with respect to $x=\pm \pi /2$ are naturally
looked for, at $|x|\neq \pi /2$, as
\begin{equation}
U(x)=U_{0}\cos \left( \sqrt{-2k}\left\vert x+\frac{\pi }{2}\right\vert
+\delta \right) ,  \label{cos}
\end{equation}%
where eigenvalue $-k$ is positive (in the presence of the repulsive
potential, energy eigenvalues of the linear Schr\"{o}dinger equation may be,
obviously, only positive), and $\delta $ is a phase shift which should be
found along with $k$. Note that, while the symmetry of expression (\ref%
{k-odd}) with respect to $x=-\pi /2$ is obvious, the symmetry with respect
to $x=+\pi /2$ follows from the fact that the coordinate takes values in
interval (\ref{x}) (i.e., formal values $x_{\mathrm{formal}}=\pi +y$, with $%
0<y<\pi $, are replaced by actual ones, $x_{\mathrm{formal}}\rightarrow x=x_{%
\mathrm{formal}}-2\pi \equiv -\left( \pi -y\right) $).

The substitution of $U(x)$ from Eq. (\ref{cos}) in jump conditions (\ref%
{jump}) leads to equations%
\begin{gather}
\tan \delta =-\frac{\varepsilon }{\sqrt{-2k}},  \label{tan} \\
\tan \left( \sqrt{-2k}\pi +\delta \right) =\frac{\varepsilon }{\sqrt{-2k}}.
\label{tan2}
\end{gather}%
Then, substituting $\delta =-\arctan \left( \varepsilon /\sqrt{-2k}\right) $%
, which follows from Eq. (\ref{tan}), in Eq. (\ref{tan2}) leads to a
transcendental equation which determines eigenvalues of the propagation
constant $k$:
\begin{equation}
\sqrt{-2k}\pi -2\arctan \left( \frac{\varepsilon }{\sqrt{-2k}}\right) =\frac{%
\pi }{2}n,\text{~}n=0,2,4,6,..,  \label{n}
\end{equation}%
where $n$ is the even quantum number, whose value $0$ corresponds to the GS,
while other values define the number of the respective even ESs$.$ In the
general case, Eq. (\ref{n}) must be solved numerically. An approximate
solution can be found in the case of weak potential barriers, $\varepsilon
\ll 1$:
\begin{equation}
k_{\mathrm{small~}\epsilon }\approx \left\{
\begin{array}{c}
-n^{2}/8-2\varepsilon /\pi ,~\mathrm{for}~n=2,4,6,...~, \\
-\varepsilon /\pi ,~\mathrm{for}~n=0,%
\end{array}%
\right.  \label{k}
\end{equation}%
as well as in the limit of very strong barriers, $\epsilon \gg 1$:%
\begin{equation}
k_{\mathrm{large~}\epsilon }\approx -\frac{1}{2}\left( \frac{n}{2}+1\right)
^{2}.  \label{n+1}
\end{equation}%
Obviously, in the case of $\epsilon =0$ eigenvalues (\ref{k}) coincide with
those for the free-space ring, $k=-n^{2}/8$, cf. Eq. (\ref{k-odd}). 0n the
other hand, in the limit of $\epsilon \rightarrow \infty $, Eq. (\ref{n+1})
demonstrates that the effect of very tall barriers amounts to the shift of
the even quantum number, $n\rightarrow n+2$.

\begin{figure}[h]
\begin{center}
\includegraphics[height=4.2cm]{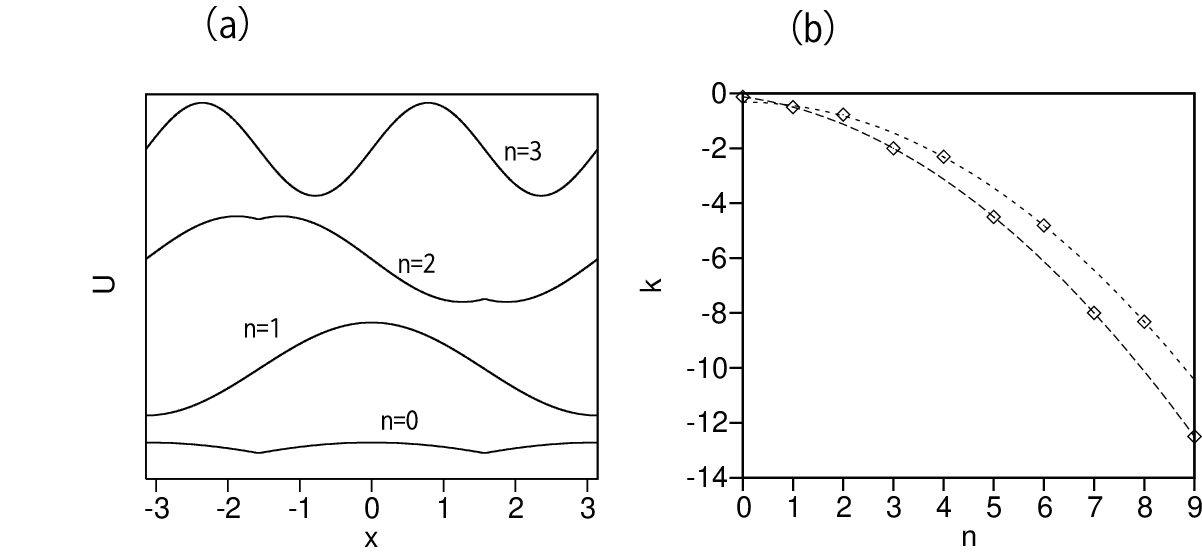}
\end{center}
\caption{(a) The GS ($n=0$) and ES ($n=1,2,3$) wavefunctions produced by
numerical solution of Eq. (\protect\ref{Ulin}) with $\protect\epsilon =0.5$.
(b) Chains of rhombuses represent numerically found eigenvalues $k$ vs.
quantum number $n$ for $\protect\epsilon =0.5$. The bottom and top dashed
lines represent, respectively, the exact analytical expression (\protect\ref%
{k-odd}) for the odd ESs and the analytical approximation (\protect\ref{k})
for the even ones (the diffence between the corrections $-2\protect%
\varepsilon /\protect\pi $ and $-\protect\varepsilon /\protect\pi $ in the
top and bottom lines of Eq. (\protect\ref{k}) is not clearly visible on the
scale of the figure).}
\label{fig1}
\end{figure}

Proceeding to the presentation of the numerical solutions of the eigenstate
problem, Fig. \ref{fig1}(a) shows examples of GS ($n=0$) and ES ($n=1$ and $%
3 $ for the odd ones, and $n=2$ for the even ES) wavefunctions produced by a
numerical solution of Eq. (\ref{Ulin}) with $\epsilon =0.5$. The numerically
found results for $\epsilon =0.5$ are further summarized in Fig. \ref{fig1}%
(b), which displays the corresponding spectrum, i.e., the relation between
eigenvalue $k$ and quantum number $n$. The exact analytical expression (\ref%
{k-odd}) and the approximate one given by the top line in Eq. (\ref{k}) are
plotted, severally, by the bottom and top dashed curves in Fig. \ref{fig1}%
(b). In particular, the numerical value of the GS\ eigenvalue is $%
k_{n=0}(\epsilon =0.5)\approx -0.1254$, while approximation (\ref{k}) yields
$\left( k_{\mathrm{small~}\epsilon }\right) _{n=0}(\epsilon =0.5)=-\epsilon
/\pi \approx -0.1591$ for the same case (note that Eq. (\ref{k}) was derived
for small $\epsilon $, while $\epsilon =0.5$ is not quite small).

\begin{figure}[h]
\begin{center}
\includegraphics[height=4.2cm]{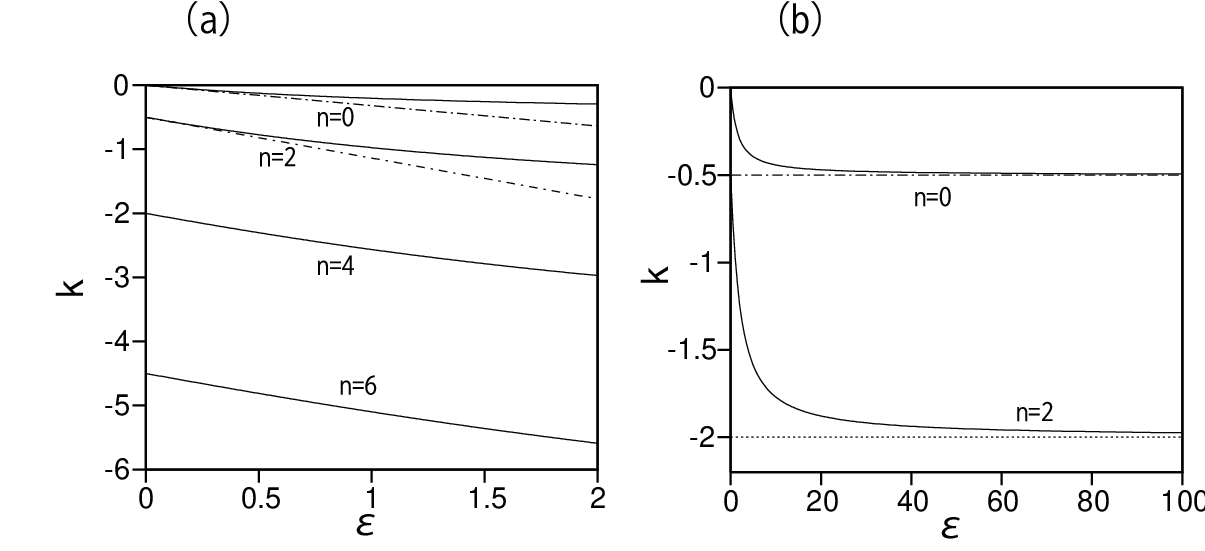}
\end{center}
\caption{(a) Eigenvalues $k$ for even $n$ vs. moderate values of the
potential-barrier height, $\protect\epsilon $, obtained from the numerical
solution of Eq.~(\protect\ref{Ulin}). The dashed lines represent the
approximation produced by Eq. (\protect\ref{k}). (b) $k$ vs. $\protect%
\epsilon $ for $n=0$ and $2$ for large values of $\protect\epsilon $. The
horizintal lines represent the respective asymptotic values (\protect\ref%
{n+1}).}
\label{fig2}
\end{figure}

For larger and very large values of $\epsilon $, the numerically found
eigenvalues and their comparison with the analytical approximations given by
Eqs. (\ref{k}) and (\ref{n+1}) are presented in Figs. \ref{fig2}(a) and (b),
for $n=0,2,4,6$ and $n=0,2$, respectively. In particular. Figs. \ref{fig2}%
(b) demonstrates slow convergence to the asymptotic values (\ref{n+1}) with
the growth of $\epsilon $.

\section{Spontaneous symmetry breaking (SSB) in the case of self-focusing ($%
\protect\sigma =+1$)}

\subsection{An example: a particular exact solution}

An exact solution of Eq. (\ref{U}) with $\sigma =+1$, which demonstrates
strongly broken symmetry, can be sought for as a \textquotedblleft
curtailed" bright soliton in the half-ring, $-\pi /2<x<+\pi /2$:%
\begin{equation}
U_{\mathrm{sol}}(x)=\sqrt{2k}\mathrm{sech}\left( \sqrt{2k}x\right) ,
\label{sol}
\end{equation}%
and a constant solution in the other half-ring, $-\pi <x<-\pi /2$ and $+\pi
/2<x<+\pi $:%
\begin{equation}
U_{\mathrm{const}}\equiv \sqrt{k}.  \label{const}
\end{equation}%
The conditions of the continuity at $x=\pm \pi /2$, i.e., $U_{\mathrm{sol}%
}(x=\pm \pi /2)=U_{\mathrm{const}}\equiv \sqrt{k}$, along with the jump
conditions (\ref{jump}) demonstrate, after a simple algebra, that such an
exact solution exists at a single value of $\epsilon $, and at single value
of $k$:%
\begin{equation}
\epsilon =\frac{1}{\sqrt{2}\pi }\ln \left( \sqrt{2}+1\right) \approx
0.198,~k=4\epsilon ^{2}\approx 0.157.  \label{eps-k}
\end{equation}

\begin{figure}[h]
\begin{center}
\includegraphics[height=4.2cm]{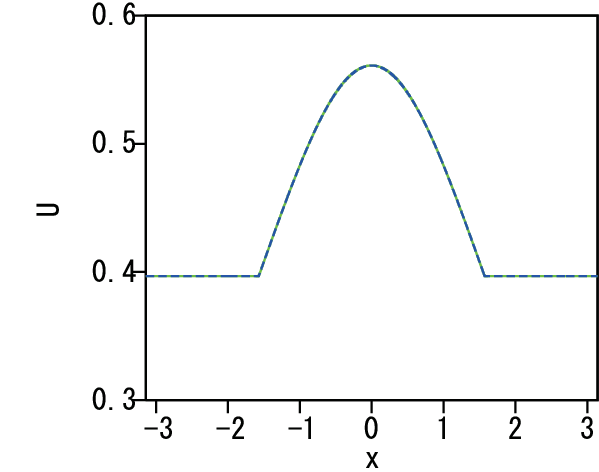}
\end{center}
\caption{The stable exact solution, given by expressions (\protect\ref{sol})
and (\protect\ref{const}), with parameters taken as per Eq. (\protect\ref%
{eps-k}), which demonstrates strong dissimilarity between the two
half-circles. The dashed blue and solid green lines represent the analytical
solution per se, and its counterpart produced by the numerical solution of
Eq. (\protect\ref{U}), with the delta-functions replaced by the Gaussian
approximation.}
\label{fig10}
\end{figure}

It is relevant to calculate the norms of the two parts of the exact
solution:
\begin{eqnarray}
N_{\mathrm{sol}} &=&\frac{2\sqrt{2}}{\pi }\ln \left( \sqrt{2}+1\right)
\approx 0.793,  \label{Nsol} \\
N_{\mathrm{const}} &=&\frac{2}{\pi }\ln ^{2}\left( \sqrt{2}+1\right) \approx
0.495,  \label{Nconst}
\end{eqnarray}%
with the ratio between them $N_{\mathrm{sol}}/N_{\mathrm{const}}=\sqrt{2}%
/\ln \left( \sqrt{2}+1\right) \approx \allowbreak 1.\,\allowbreak 602$. This
stationary solution is plotted in Fig. \ref{fig10}. Direct simulations of
Eq. (\ref{NLS}) for the perturbed propagation of this stationary state
demonstrate that it is \emph{stable} (not shown here in detail).

\subsection{The variational approximation (VA)}

\subsubsection{VA for stationary states: the prediction of SSB}

Solutions of the full stationary nonlinear equation (\ref{U}) are
characterized by their total norm and energy:%
\begin{equation}
N=N_{1}=\int_{-\pi /2}^{+\pi /2}U^{2}(x)dx+\left( \int_{-\pi }^{-\pi
/2}+\int_{+\pi /2}^{+\pi }\right) U^{2}(x)dx\equiv N_{1}+N_{2},  \label{N}
\end{equation}%
\begin{equation}
E=\int_{-\pi }^{+\pi }\left[ \frac{1}{2}\left( \frac{dU}{dx}\right) ^{2}-%
\frac{\sigma }{2}U^{4}\right] dx+\epsilon \left[ U^{2}\left( x=+\frac{\pi }{2%
}\right) +U^{2}\left( x=-\frac{\pi }{2}\right) \right] .  \label{E}
\end{equation}%
To seek for stationary states of the nonlinear system, we first, apply VA,
based on the following ansatz:%
\begin{equation}
U(x)=\left\{
\begin{array}{c}
a_{0}+a_{1}\cos x,\;\;\;\mathrm{for}\;|x|<\pi /2, \\
a_{0}-a_{2}\cos x,\;\;\;\mathrm{for}\;\pi /2<|x|<\pi .%
\end{array}%
\right.  \label{ans}
\end{equation}%
The substitution of the ansatz in the jump conditions (\ref{jump}) leads to
a relation between the amplitudes,%
\begin{equation}
a_{1}+a_{2}=2\epsilon a_{0}.  \label{120}
\end{equation}%
Further, the ansatz gives rise to the following expressions for norms $N_{1}$
and $N_{2}$ of the two layers, defined as per Eq. (\ref{N})
\begin{eqnarray}
N_{1} &=&\pi a_{0}^{2}+\pi a_{1}^{2}/2+4a_{0}a_{1},  \notag \\
N_{2} &=&\pi a_{0}^{2}+\pi a_{2}^{2}/2+4a_{0}a_{2}.  \label{12}
\end{eqnarray}%
The energy (\ref{E}) corresponding to ansatz (\ref{ans}) is calculated
analytically as
\begin{eqnarray}
E &=&\frac{\pi }{4}(a_{1}^{2}+a_{2}^{2})-\pi \left[ a_{0}^{4}+\frac{3}{2}%
a_{0}^{2}\left( a_{1}^{2}+a_{2}^{2}\right) +\frac{3}{16}\left(
a_{1}^{4}+a_{2}^{4}\right) \right]  \notag \\
&&-4a_{0}^{3}\left( a_{1}+a_{2}\right) -\frac{8}{3}a_{0}\left(
a_{1}^{3}+a_{2}^{3}\right) +2\epsilon a_{0}^{2}.  \label{Eans}
\end{eqnarray}

In terms of ansatz (\ref{ans}), the symmetric state corresponds to $%
a_{1}=a_{2}$. In view of the relation (\ref{120}) between the amplitudes of
the ansatz, SSB may be sought for by setting
\begin{equation}
a_{1}=\epsilon a_{0}+B/2,~a_{2}=\epsilon a_{0}-B/2,  \label{a12}
\end{equation}%
the onset of SSB implying a transition to a solution with $B\neq 0$. In
terms of amplitudes $a_{0}$ and $B$, energy (\ref{Eans}) $E$ is rewritten as%
\begin{gather}
E=\frac{\pi }{8}B^{2}-\frac{3}{128}\pi B^{4}+\left( \frac{\pi }{2}\epsilon
^{2}+2\epsilon \right) a_{0}^{2}  \notag \\
-\left[ \pi \left( 1+3\epsilon ^{2}+\frac{3}{8}\epsilon ^{4}\right)
+8\epsilon +\frac{16}{3}\epsilon ^{2}\right] a_{0}^{4}  \notag \\
-\left[ \pi \left( \frac{3}{4}+\frac{9}{16}\epsilon ^{2}\right) +4\epsilon %
\right] a_{0}^{2}B^{2}.  \label{E2}
\end{gather}

By the substitution of expressions (\ref{a12}) in the definition of the
norm, based on Eqs. (\ref{N}) and (\ref{12}), one can express the basic
amplitude $a_{0}$ in terms of the total norm, $N$, and SSB amplitude $B$:%
\begin{equation}
a_{0}^{2}=\frac{N-\pi B^{2}/4}{2\pi +8\epsilon +\pi \epsilon ^{2}}.
\label{a0}
\end{equation}%
Lastly, with the use of this relation, energy (\ref{E2}) is cast in the form
of a cumbersome but tractable expression, written in terms of $B$ and $N$:
\begin{equation}
E=c_{0}+(c_{20}-c_{21}N)B^{2}+c_{4}B^{4},  \label{E3}
\end{equation}%
where
\begin{equation}
c_{0}=\frac{(\pi \epsilon ^{2}/2+2\epsilon )N}{2\pi +8\epsilon +\pi \epsilon
^{2}}-\frac{\left[ \pi (1+3\epsilon ^{2}+3\epsilon ^{4}/8)+8\epsilon
+16\epsilon ^{3}/3\right] N^{2}}{(2\pi +8\epsilon +\pi \epsilon ^{2})^{2}},
\label{c0}
\end{equation}%
\begin{equation}
c_{20}=\frac{\pi ^{3}/2+3\pi ^{2}\epsilon +(\pi ^{3}/4+4\pi )\epsilon
^{2}+\pi ^{2}\epsilon ^{3}/2}{(2\pi +8\epsilon +\pi \epsilon ^{2})^{2}},
\label{c20}
\end{equation}%
\begin{equation}
c_{21}=\frac{\pi ^{2}+10\pi \epsilon +(32+3\pi ^{2}/8)\epsilon ^{2}+35\pi
\epsilon ^{3}/6+3\pi ^{2}\epsilon ^{4}/8}{(2\pi +8\epsilon +\pi \epsilon
^{2})^{2}},  \label{c21}
\end{equation}%
\begin{equation}
c_{4}=\frac{7\pi ^{3}/32+9\pi ^{2}\epsilon /4+(13\pi /2+3\pi
^{3}/16)\epsilon ^{2}+17\pi ^{2}\epsilon ^{3}/12+3\pi ^{3}\epsilon ^{4}/32}{%
(2\pi +8\epsilon +\pi \epsilon ^{2})^{2}}.  \label{c4}
\end{equation}

The variational principle states that, for a given norm and fixed $\epsilon $%
, the system's GS is predicted as one with a positive value of $B^{2}$ which
provides for a minimum of energy (\ref{E3}), i.e.,%
\begin{equation}
B^{2}=\frac{c_{21}N-c_{20}}{2c_{4}}.  \label{B}
\end{equation}%
Because coefficients $c_{20}$, $c_{21}$, and $c_{4}$ are defined by
expressions (\ref{c20}), (\ref{c21}), and (\ref{c4}) as positive ones, Eq. (%
\ref{B}) predicts that SSB commences when the norm exceeds a critical value,%
\begin{equation}
N_{\mathrm{c}}(\epsilon )=c_{20}/c_{21}.  \label{Nc}
\end{equation}

Further, general properties of the energy (Hamiltonian) imply that the
eigenvalue corresponding to the present solution can be calculated as $%
k=-\partial E/\partial N$. The respective VA-produced result for the
symmetric state, with $B=0$ in Eq. (\ref{a12}), is%
\begin{equation}
k=k_{\mathrm{s}}\equiv -\frac{\partial c_{0}}{\partial N}=-\frac{(\pi
\epsilon ^{2}/2+2\epsilon )}{2\pi +8\epsilon +\pi \epsilon ^{2}}+2N\frac{%
\left[ \pi (1+3\epsilon ^{2}+3\epsilon ^{4}/8)+8\epsilon +16\epsilon ^{3}/3%
\right] }{(2\pi +8\epsilon +\pi \epsilon ^{2})^{2}}.  \label{ksymm}
\end{equation}%
while for the broken-symmetry state it is
\begin{equation}
k=k_{\mathrm{as}}=k_{s}+\frac{(c_{21}N-c_{20})c_{21}}{(2c_{4})}.
\label{kasymm}
\end{equation}

\subsubsection{VA for nonstationary states: the stability analysis}

The VA may also be applied as an approximation predicting the evolution of
nonstationary ($z$-dependent) states (in particular, the onset of MI) \cite%
{Anderson,progress}. To this end, we use the Lagrangian of the full equation
(\ref{NLS}), in its nonstationary form:
\begin{eqnarray}
L &=&\frac{1}{2}\int_{-\pi }^{+\pi }\left[ i\left( \frac{\partial u}{%
\partial z}u^{\ast }-\frac{\partial u^{\ast }}{\partial z}u\right)
-\left\vert \frac{\partial u}{\partial x}\right\vert ^{2}+|u|^{4}\right] dx
\notag \\
&&-\epsilon \left[ \left\vert u\left( x=+\frac{\pi }{2}\right)
^{2}\right\vert +\left\vert u\left( x=-\frac{\pi }{2}\right) ^{2}\right\vert %
\right] .  \label{Lagr}
\end{eqnarray}%
To derive the dynamical version of the VA, we adopt the following\textit{\
ans\"{a}tze} for the symmetric and asymmetric states:%
\begin{eqnarray}
u_{\mathrm{s}}^{\mathrm{(VA)}}\left( x,z\right) &=&a_{0}\left[ 1+\epsilon
|\cos x|\right] \exp (ikz),  \label{us} \\
u_{\mathrm{as}}^{\mathrm{(VA)}}\left( x,z\right) &=&\left[ a_{0}+\epsilon
a_{0}|\cos x|+b(z)\cos x\right] \exp (ikt),  \label{uas}
\end{eqnarray}%
where $a_{0}$ and $k$ are real constants, while $b(z)$ is a complex variable
amplitude which accounts for the onset of \textit{spatial modulation} of the
symmetric state, or, in other words, for the initiation of SSB. Note that
the norm (\ref{N}) corresponding to the symmetric ansatz (\ref{us}) is%
\begin{equation}
N_{\mathrm{s}}=a_{0}^{2}\left( 2\pi +8\epsilon +\pi \epsilon ^{2}\right) .
\label{Ns}
\end{equation}%
Naturally, the expression for $U(x)\equiv \left\vert u_{\mathrm{s}}^{\mathrm{%
(VA)}}\left( x,z\right) \right\vert $, as given by Eq. (\ref{us}), is
identical to ansatz (\ref{ans}) adopted above for stationary solutions, in
the case of the symmetric state, as it follows from Eq. (\ref{a12}) with $%
B=0 $.

The substitution of ansatz (\ref{uas}) in Lagrangian (\ref{Lagr}) yield the
corresponding VA Lagrangian,
\begin{gather}
L_{\mathrm{VA}}=i\frac{\pi }{2}\left( \frac{db}{dz}b^{\ast }-\frac{db^{\ast }%
}{dz}b\right) -\pi k|b|^{2}-\frac{\pi }{2}|b|^{2}  \notag \\
+\left[ \frac{1}{2}a_{0}^{2}(\left( b^{\ast }\right)
^{2}+b^{2})+2a_{0}^{2}|b|^{2}\right] \left( \pi +\frac{16}{3}\epsilon +\frac{%
3}{4}\pi \epsilon ^{2}\right) +\frac{3}{8}\pi |b|^{4},  \label{LVA}
\end{gather}%
with $\ast $ standing for the complex conjugate. Then, the evolution
equation for $b(z)$ is derived from the VA Lagrangian (\ref{LVA}) as the
corresponding Euler-Lagrange (EL) equation, \textit{viz}.,
\begin{equation}
i\frac{db}{dz}=\left( k+\frac{1}{2}\right) b-\left( 1+\frac{16}{3\pi }%
\epsilon +\frac{3}{4}\epsilon ^{2}\right) a_{0}^{2}\left( b^{\ast
}+2b\right) -\frac{3}{4}|b|^{2}b.  \label{EL}
\end{equation}%
The onset of the instability is signaled by the emergence of an
exponentially growing solution of the linearized version of Eq. (\ref{EL})
for $b(z)$. To analyze this scenario, the linearized equation is split in
real and imaginary parts, by substituting $b(z)\equiv b_{1}(z)+ib_{2}(z)$:%
\begin{gather}
\frac{db_{1}}{dz}=\left[ \left( k+\frac{1}{2}\right) -a_{0}^{2}\left( 1+%
\frac{16}{3\pi }\epsilon +\frac{3}{4}\epsilon ^{2}\right) \right] b_{2},
\label{b1} \\
-\frac{db_{2}}{dz}=\left[ \left( k+\frac{1}{2}\right) -3a_{0}^{2}\left( 1+%
\frac{16}{3\pi }\epsilon +\frac{3}{4}\epsilon ^{2}\right) \right] b_{1}.
\label{b2}
\end{gather}%
Solutions to this system of linear equations are looked for as $%
b_{1,2}(z)=b_{1,2}^{(0)}\exp \left( \lambda z\right) $. A straightforward
calculation yields the value of the instability growth rate $\lambda $, as
\begin{gather}
\lambda ^{2}=\pi \left[ \left( k+\frac{1}{2}\right) -a_{0}^{2}\left( 1+\frac{%
16}{3\pi }\epsilon +\frac{3}{4}\epsilon ^{2}\right) \right]  \notag \\
\times \left[ 3a_{0}^{2}\left( 1+\frac{16}{3\pi }\epsilon +\frac{3}{4}%
\epsilon ^{2}\right) -\left( k+\frac{1}{2}\right) \right] .  \label{lambda}
\end{gather}

In the framework of the present approximation, MI takes place if Eq. (\ref%
{lambda}) yields $\allowbreak \lambda ^{2}>0$. As it follows from Eq. (\ref%
{lambda}) and expressions (\ref{ksymm}) and (\ref{Ns}) for the propagation
constant and norm of the symmetric state, this condition eventually amounts
to
\begin{equation}
N>N_{\mathrm{c}}=\frac{\pi \left[ 2\pi ^{2}+12\pi \epsilon +(\pi
^{2}+16)\epsilon ^{2}+2\epsilon ^{3}\right] }{4\pi ^{2}+40\pi \epsilon
+(3\pi ^{2}/2+128)\epsilon ^{2}+70\pi \epsilon ^{3}/3+3\pi ^{2}\epsilon
^{4}/2}.  \label{Ncc}
\end{equation}%
Note that the critical value of the norm, given by Eq. (\ref{Ncc}), \emph{is
identical} to one for the onset of SSB, which was found above in the form of
Eqs. (\ref{Nc}) and (\ref{c20}), (\ref{c21}), hence the MI is precisely the
instability mode of the symmetric state which determines the commencement of
the SSB.

The comparison of the VA predictions with numerical results is provided
below in Figs. \ref{fig3}, \ref{fig4}, and \ref{fig6}(a).

\subsection{The analytical approximation for large $\protect\epsilon $}

Another approximation can be elaborated in the limit of large $\epsilon $,
when the ring is split by the delta-functional barriers in two half-rings,
which are nearly isolated from each other. The weak residual coupling
between the two half-rings implies that weak self-focusing nonlinearity,
measured by the squared amplitude of the solution, $U_{0}^{2}$ (see Eqs. (%
\ref{cos})), is sufficient to break the symmetry. This means that, in the
zero-order approximation, in the half-ring $-\pi /2<x<+\pi /2$ one may take
the obvious solutions of the linearized equation (\ref{Ulin}),%
\begin{equation}
U(x)=U_{0}\cos \left( \sqrt{-2k}x\right) .  \label{U0}
\end{equation}%
The first nonlinear correction to this solution follows from the elementary
formula
\begin{equation}
\cos ^{3}\left( \sqrt{-2k}x\right) =(3/4)\cos \left( \sqrt{-2k}x\right)
+(1/4)\cos \left( 3\sqrt{-2k}x\right) .  \label{cubic}
\end{equation}%
The former term on the right-hand side of Eq. (\ref{cubic}) implies that $k$
in Eq. (\ref{U}) with $\sigma =+1$ is replaced by the corrected value,%
\begin{equation}
k\rightarrow k-(3/4)U_{0}^{2},  \label{shift}
\end{equation}%
and the third-harmonic correction added to the fundamental one (\ref{U0}) is%
\begin{equation}
U_{3}(x)=-(32k)^{-1}U_{0}^{3}\cos \left( 3\sqrt{-2k}x\right) ,  \label{U3}
\end{equation}

In the limit of $\epsilon =\infty $, the boundary conditions for GS (\ref{U0}%
), become $U\left( x=\pm \pi /2\right) $ $=0$, implying $k=-1/2$. At large
but finite $\epsilon $, a small correction is introduced:
\begin{equation}
k=-1/2+\delta k,~\mathrm{i.e.},~\sqrt{-2k}\approx 1-\delta k.  \label{delta}
\end{equation}%
On the other hand, the contribution of correction (\ref{U3}) to the boundary
conditions (\ref{jump}) is negligible in the first approximation because $%
\cos \left( 3x\right) =0$ at $x=\pm \pi /2$. The only essential correction
originating from the weak nonlinearity is produced by the shift (\ref{shift}%
) of the propagation constant, which replaces $U_{0}\cos \left( \sqrt{-2k}%
x\right) $ in Eq. (\ref{U0}) by%
\begin{equation}
U(x)\approx U_{0}\cos \left( \left( 1-\delta k+(3/4)U_{0}^{2}\right)
x\right) ,  \label{corrected}
\end{equation}%
where expression (\ref{delta}) is substituted for $\sqrt{-2k}$.

The GS solution with broken symmetry is approximated by expressions (\ref%
{corrected}) in the two half-rings, with different amplitudes, $%
U_{0}\rightarrow U_{1,2}$. The substitution of these expressions in the
condition of the continuity of $U(x)$ at $x=\pm \pi /2$ yields the following
equation, in the lowest approximation with respect to small parameters $%
\delta k$ and $U_{1,2}^{2}$:%
\begin{equation}
U_{1}\left[ \delta k-(3/4)U_{1}^{2}\right] =U_{2}\left[ \delta
k-(3/4)U_{2}^{2}\right]   \label{cont}
\end{equation}%
Further, the substitution of expressions (\ref{corrected}) with amplitudes $%
U_{1,2}$ in the jump condition (\ref{jump}) leads to the following equation,
which is also written in the lowest approximation with respect to small
parameters $\delta k$, $U_{1,2}^{2}$, and $1/\epsilon $:%
\begin{equation}
\frac{2}{\pi \epsilon }\left( U_{1}+U_{2}\right) =\delta k\cdot \left(
U_{1}+U_{2}\right) -(3/4)\left( U_{1}^{3}+U_{2}^{3}\right) .
\label{cos-jump}
\end{equation}%
When writing Eq. (\ref{cos-jump}), the right-hand side in Eq. (\ref{jump})
is taken in the symmetrized form, as the half-sum of $U(x=\pm \pi /2)$
corresponding to the two half-rings (because of the continuity condition,
the half-sum is the same as $U(x=\pm \pi /2)$ corresponding to either
half-ring, the use of the symmetrized form making it easier to solve the
equations).

For the symmetry-breaking GS, with $U_{1}\neq U_{2}$, one can cancel factor $%
\left( U_{1}-U_{2}\right) $ in the continuity equation (\ref{cont}), and
cancel factor $\left( U_{1}+U_{2}\right) $ in the jump equation (\ref%
{cos-jump}), arriving at the following equations:%
\begin{equation}
U_{1}^{2}+U_{2}^{2}+U_{1}U_{2}=(4/3)\delta k,  \label{+}
\end{equation}%
\begin{equation}
U_{1}^{2}+U_{2}^{2}-U_{1}U_{2}=(4/3)\left[ \delta k-2/\left( \pi \epsilon
\right) \right] .  \label{-}
\end{equation}%
The combination of Eqs. (\ref{+}) and (\ref{-}) give rise to a simple
relation,
\begin{equation}
U_{1}U_{2}=4/\left( 3\pi \epsilon \right) .  \label{U1U2}
\end{equation}
Using it to eliminate $U_{2}$ in favor of $U_{1}$, one arrives at the
biquadratic equation for $U_{1}$:%
\begin{equation}
U_{1}^{4}-\frac{4}{3}\left( \delta k-\frac{1}{\pi \epsilon }\right)
U_{1}^{2}+\left( \frac{4}{3\pi \epsilon }\right) ^{2}=0.  \label{quadr}
\end{equation}%
The result following from Eq. (\ref{quadr}) is that the symmetry-breaking GS
state exists under the following threshold condition,%
\begin{equation}
\delta k>\left( \delta k\right) _{\mathrm{thr}}\equiv 3/\left( \pi \epsilon
\right) .  \label{thr}
\end{equation}%
The amplitude at the threshold point is found by substituting $%
U_{1}=U_{2}\equiv U_{\mathrm{thr}}$ in Eq. (\ref{U1U2}), which yields $U_{%
\mathrm{thr}}^{2}=4/\left( 3\pi \epsilon \right) $. The respective critical
value of the norm, which determines the onset of SSB at large $\epsilon $, is%
\begin{equation}
N_{\mathrm{c}}=\pi U_{\mathrm{thr}}^{2}\equiv 4/(3\epsilon )  \label{Nthr}
\end{equation}%
(cf. the prediction (\ref{Ncc}) provided by the VA). As shown below in Fig. %
\ref{fig4}, this approximation is accurate enough at very large values of $%
\epsilon $.

\subsection{Numerical findings}

\begin{figure}[h]
\begin{center}
\includegraphics[height=3.7cm]{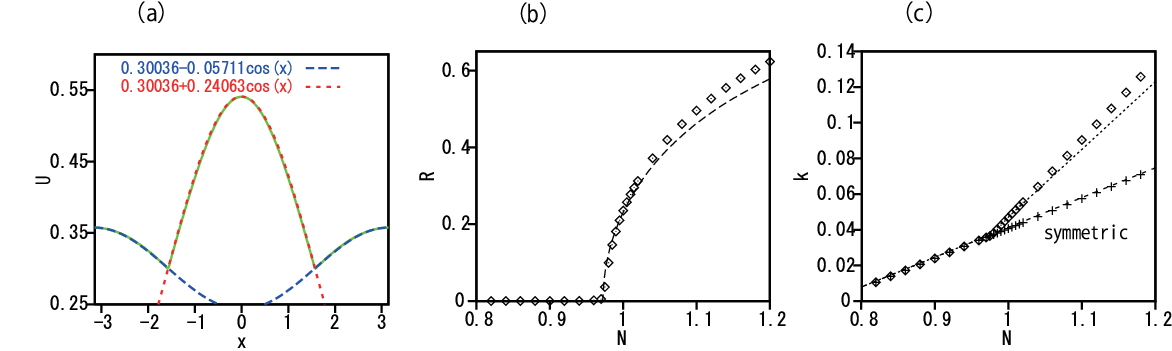}
\end{center}
\caption{(a) The solid green lines represent the numerically obtained
profile of the stable asymmetric mode $U(x)$ with $\protect\sigma =+1$
(self-focusing nonlinearity), $\protect\epsilon =0.5$ and $N=1.02$. The
corresponding ansatz (\protect\ref{ans}) with coefficients (\protect\ref{012}%
) predicted by the VA (and printed in the panel), is plotted by the dashed
blue and red lines. (b) The numerically found value of asymmetry measure (%
\protect\ref{R}) vs. the total norm, $N$, at $\protect\epsilon =0.5$. The
dashed line is the VA prediction for $R$, see Eqs. (\protect\ref{12}), (%
\protect\ref{a12}), and (\protect\ref{B}). (c) Numerically obtained values
of the propagation constant $k$ for the symmetric (pluses) and
symmetry-breaking (rhombuses) modes vs. $N$. The dashed lines represent the
respective VA prediction, see Eqs. (\protect\ref{ksymm}) and (\protect\ref%
{kasymm}).}
\label{fig3}
\end{figure}

Stationary solutions of Eq. (\ref{NLS}) with $\sigma =+1$ were produced by
means of the well-known \cite{im-time} imaginary-time-evolution method. A
typical example of a numerically found stable stationary mode $U(x)$ with
the broken symmetry, i.e., a result of SSB, is plotted in Fig. \ref{fig3}(a)
for $\epsilon =0.5$ and $N=1.02$. The corresponding variational ansatz (\ref%
{ans}) with the VA-produced coefficients, \textit{viz}.,
\begin{equation}
a_{0}=0.3004,a_{1}=0.2406,a_{2}=0.0571,  \label{012}
\end{equation}
is plotted by the dashed lines.

As a measure of the symmetry breaking, we define%
\begin{equation}
R=(N_{1}-N_{2})/(N_{1}+N_{2}),  \label{R}
\end{equation}%
with $N_{1,2}$ defined in Eq. (\ref{N}). Figure \ref{fig3}(b) summarizes the
numerical results obtained for $\epsilon =0.5$, in the form of the
bifurcation diagram represented by the dependence $R(N)$ (the chain of
rhombuses), while the dashed line represents the VA prediction for the same
case, as produced by Eqs. (\ref{12}), (\ref{a12}), and (\ref{B}). It is seen
that the SSB bifurcation is of the forward (supercritical \cite{Iooss})
type. Another SSB\ characteristic, \textit{viz}., dependences $k(N)$ for the
symmetric and asymmetric states, and their comparison to the VA prediction,
produced by Eqs. (\ref{ksymm}) and (\ref{kasymm}), is shown in Fig. \ref%
{fig3}(c).

\begin{figure}[h]
\begin{center}
\includegraphics[height=4.cm]{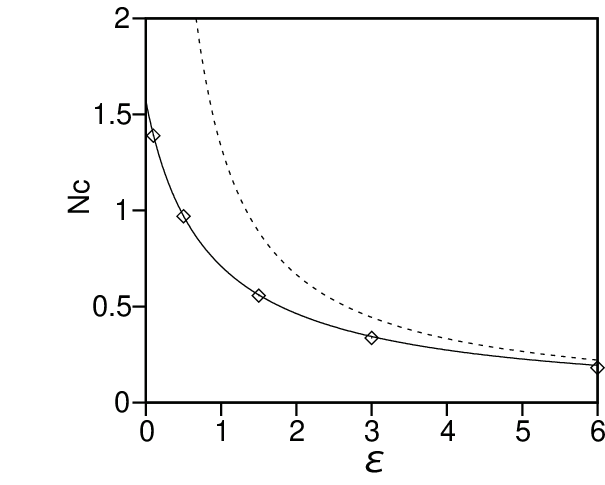}
\end{center}
\caption{The continuous curve shows the critical value $N_{\mathrm{c}}$ at
the SSB point in the model with self-focusing ($\protect\sigma =+1$), as
predicted by the VA, see Eq. (\protect\ref{Ncc}). The prediction of the
approximation for large $\protect\epsilon $, produced by Eq. (\protect\ref%
{Nthr}), is plotted by the dashed curve. Rhombuses represent numerically
found values of $N_{\mathrm{c}}$.}
\label{fig4}
\end{figure}

The results are further summarized in Fig. \ref{fig4}, which reports
numerically found values of the critical norm at the SSB point, $N_{\mathrm{c%
}}$, for $\epsilon =0.1$, $0.5$, $1.5$, and $3$, along with the curve
plotting the VA prediction (\ref{Ncc}) for the $N_{\mathrm{c}}(\epsilon )$
dependence. Note that the value $N_{\mathrm{c}}(\epsilon =0)=\pi /2$
corresponds to the commonly known threshold of MI (alias the Benjamin-Feir
instability \cite{Stuart,Efim}) for the uniform (CW) state in the ring \cite%
{Agrawal}. Figures \ref{fig3} and \ref{fig4} demonstrate high accuracy of
the VA, in comparison to the numerical findings. As mentioned above, the
decrease of $N_{\mathrm{c}}$ with the increase of $\epsilon $ is a natural
property of the system, as larger $\epsilon $ means weaker coupling between
two semi-rings, hence weaker nonlinearity is sufficient to initiate SSB.

\begin{figure}[h]
\begin{center}
\includegraphics[height=4.2cm]{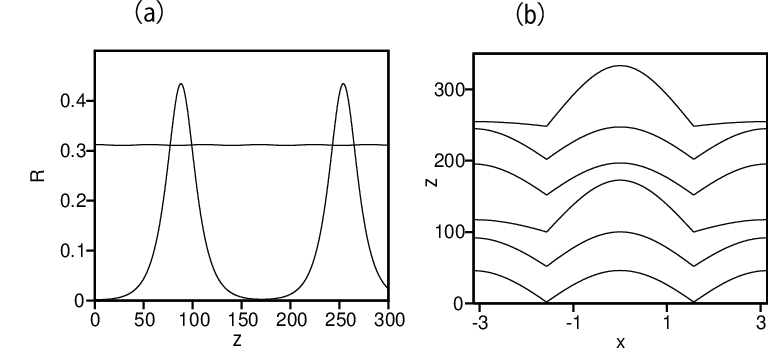}
\end{center}
\caption{(a) The evolution of the asymmetry measure $R(z)$ (see Eq. (\protect
\ref{R})) for the unstable symmetric and stable asymmetric states in the
model with the self-focusing nonlinearity ($\protect\sigma =+1$), which
coexist at $\protect\epsilon =0.5$ and $N=1.02$. (b) Seven snapshots of $%
|u(x)|$ at $z=50n$ $(n=0,1,2\cdots ,6)$, starting from the symmetric
solution with a very small perturbation added to it.}
\label{fig5}
\end{figure}

The stability of the symmetric and asymmetric stationary states was verified
by direct simulations of their perturbed evolution in the framework of Eq.~(%
\ref{NLS}). An example is presented in Fig. \ref{fig5} for $\epsilon =0.5$
and $N=1.02$. Figure \ref{fig5}(a) shows the corresponding evolution of the
asymmetry parameter $R$, defined as per Eq. (\ref{R}), It is seen that $R$
remains practically constant for the stable asymmetric state. On the other
hand, the instability of the symmetric one leads to large oscillations of $R$
between $0$ and $0.435$, i.e., periodic oscillations between symmetric and
strongly asymmetric configurations. The oscillations are illustrated in Fig. %
\ref{fig5}(b) by means of $7$ snapshots of $|u(x)|$, starting from the
initial symmetric configuration at $z=0$ and periodically attaining a fully
asymmetric one (e.g., at $z=250$).

\begin{figure}[h]
\begin{center}
\includegraphics[height=4.2cm]{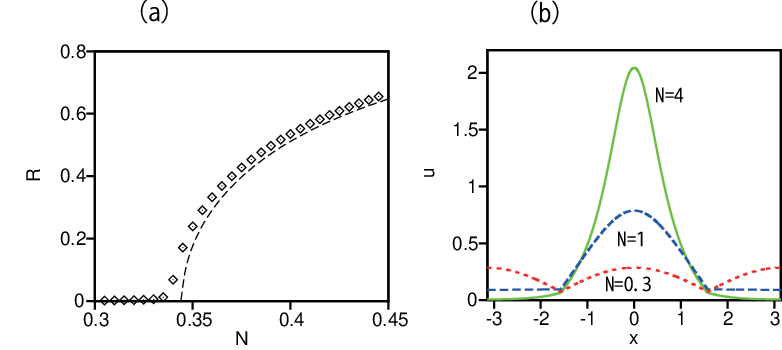}
\end{center}
\caption{(a) The chain of rhombuses presents the dependence of the asymmetry
measure (\protect\ref{R}) on the total norm, $N$, as produced by the
numerical solution of Eq. (\protect\ref{U}) with $\protect\sigma =+1$
(self-focusing) and $\protect\epsilon =3$.\ The dashed line shows the same
dependence, as predicted by the VA, see Eq. (\protect\ref{Ncc}). (b)
Numerically found profiles of $U(x)$, with $\protect\epsilon =3$,\ for a
stable symmetric mode wtih $N=0.3$ (the red dotted line), and stable
asymmetric ones, with $N=1$ and $4$ (the blue dashed and green solid lines,
respectively).}
\label{fig6}
\end{figure}

While the above examples, presented in Figs. \ref{fig3} and \ref{fig5},
display the numerical and VA-predicted analytical results for the system
with relatively weak potential barriers ($\epsilon =0.5$ in Eq. (\ref{NLS}%
)), the findings obtained for much stronger barriers, with $\epsilon =3$,
are presented in Fig. \ref{fig6}. In particular, Fig. \ref{fig6}(a)
demonstrates that the VA provides a sufficiently accurate approximation for
the SSB diagram (the $R(N)$ curve) in this case too.

\section{Spontaneous antisymmetry breaking (SASB) in the case of
self-defocusing ($\protect\sigma =-1$)}

\subsection{The variational approximation (VA)}

\subsubsection{VA for stationary antisymmetric states}

As mentioned above, in the case of the self-repulsive nonlinearity, which
corresponds to $\sigma =-1$ in Eq. (\ref{NLS}), one cannot expect a
transition of the GS into a symmetry-breaking shape, but destabilization of
the spatially odd (antisymmetric) lowest ES is possible \cite{Michal,Raymond}%
. In the framework of VA, a solution of Eq. (\ref{NLS}), with $\sigma =-1$,
for the lowest ES may be approximated by the ansatz which is antisymmetric
with respect to points $x=\pm \pi /2$, where the delta-functional barriers
are set:%
\begin{equation}
u_{\mathrm{anti}}^{\mathrm{(VA)}}(x)=\left[ A_{1}\cos x+A_{3}\cos (3x)\right]
\exp (ikz).  \label{anti}
\end{equation}%
cf. ansatz (\ref{us}) for the spatially even GS. The norm of this ansatz is
\begin{equation}
N_{\mathrm{anti}}=\pi \left( A_{1}^{2}+A_{3}^{2}\right) ,  \label{Nanti}
\end{equation}

\begin{figure}[h]
\begin{center}
\includegraphics[height=4.2cm]{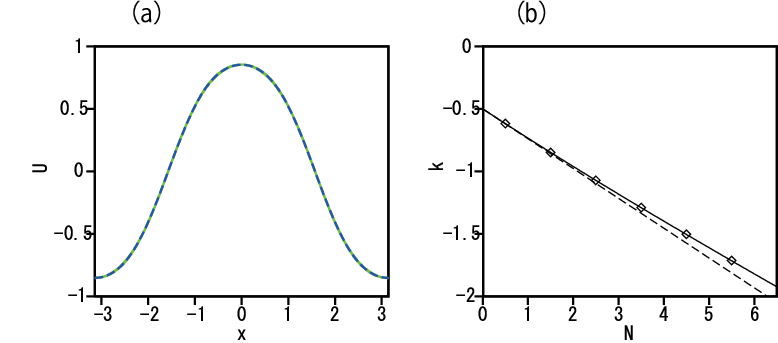}
\end{center}
\caption{(a) The profile of the stable lowest excited state (ES), found as
the numerical solution of Eq. (\protect\ref{U}) with $\protect\sigma =-1$, $%
\protect\epsilon =2$, and the total norm $N=2.5$, is plotted by the solid
green line. The profile is antisymmetric (odd) with respect to points $x=\pm
\protect\pi /2$, where the delta-functional potential barriers are set. The
dashed blue line (which actually completely overlaps with the solid green
one) represents the VA-predicted counterpart of this state, given by ansatz (%
\protect\ref{anti}), with coefficients $A_{1}=0.8912$ and $A_{3}=-0.0383$.
(b) The $k(N)$ dependence for the lowest-ES solution, as predicted by the VA
based on ansatz (\protect\ref{anti}) (the solid line), and its simplified
version with $A_{3}=0$ (the dashed line). The chain of rhombuses shows
values of $k(N)$ obtained from the numerical solution.}
\label{fig7}
\end{figure}

The substitution of the ansatz in Lagrangian (\ref{Lagr}) (with the opposite
sign in front of $|u|^{4}$, which corresponds to $\sigma =-1$), readily
produces the effective VA Lagrangian,%
\begin{equation}
\left( L_{\mathrm{VA}}\right) _{\mathrm{anti}}=-\pi (A_{1}^{2}+A_{3}^{2})k-%
\frac{\pi }{2}(A_{1}^{2}+9A_{3}^{2})-\frac{3\pi }{8}%
(A_{1}^{4}+A_{3}^{4}+4A_{1}^{2}A_{3}^{2})+\frac{\pi }{2}A_{1}^{3}A_{3}.
\label{LVAanti}
\end{equation}%
Note that this expression does not include parameter $\epsilon $, as ansatz (%
\ref{anti}) vanishes at the points where the delta-functional barriers are
set. Then, parameters of the VA-approximated odd mode are determined by the
EL equations, $\partial \left( L_{\mathrm{VA}}\right) _{\mathrm{anti}%
}/\partial \left( A_{1,3}\right) =0$. Figure \ref{fig7}(a) displays an
example of the stationary state, as produced by the numerical solution of
Eq. (\ref{U}) for $N=2.5$, along with its VA-predicted counterpart, with
amplitudes $A_{1}=0.8912$ and $A_{3}=-0.00383$, as given by the EL
equations. Figure \ref{fig7}(b) represents the family of the lowest-ES
stationary solutions by means of the respective $k(N)$ dependence, as
obtained from the numerical results (the chain of rhombuses), and as
predicted by the VA, including its simplified version with $A_{3}=0$ in
ansatz (\ref{anti}), the latter one amounting to $k=-1/2-3N/(4\pi )$.

\subsubsection{Nonstationary VA: the stability analysis of the antisymmetric
state}

Similar to the VA for the nonstationary states with broken symmetry, which
was developed in the previous section, we here address the dynamics of
states breaking the antisymmetry. For this purpose, the following ansatz is
adopted, with the complex amplitude $b(z)$ of the SASB perturbation (cf. Eq.
(\ref{uas})):
\begin{equation}
u_{\mathrm{SASB}}^{\mathrm{(VA)}}u(x,z)=\left[ A_{1}\cos x+A_{3}\cos
(3x)+b(z)(1+\epsilon |\cos x|\right] e^{ikz}.  \label{uu}
\end{equation}%
The substitution of this ansatz in the Lagrangian and neglecting terms which
are quartic with respect to the perturbation amplitude, we obtain
\begin{gather}
\left( L_{\mathrm{VA}}\right) _{\mathrm{SASB}}=i\left( \pi +4\epsilon +\frac{%
\pi }{2}\epsilon ^{2}\right) \left( \frac{\partial b}{\partial z}b^{\ast }-%
\frac{\partial b^{\ast }}{\partial z}b\right)  \notag \\
-(2\pi +8\epsilon +\pi \epsilon ^{2})k|b|^{2}-\left( 2\epsilon +\frac{\pi }{2%
}\epsilon ^{2}\right) |b|^{2}  \notag \\
-\frac{A_{1}^{2}}{2}\left( \pi +\frac{16}{3}\epsilon +\frac{3\pi }{4}%
\epsilon ^{2}\right) (b^{2}+b^{\ast 2}+4|b|^{2})  \notag \\
-\frac{A_{3}^{2}}{2}\left( \pi +\frac{144}{35}\epsilon +\frac{\pi }{2}%
\epsilon ^{2}\right) (b^{2}+b^{\ast 2}+4|b|^{2})  \notag \\
-A_{1}A_{3}\left( \frac{16}{15}\epsilon +\frac{\pi }{4}\epsilon ^{2}\right)
\left( b^{2}+(b^{\ast })^{2}+4|b|^{2}\right) .  \label{LL}
\end{gather}

Further, following the pattern of the analysis developed in the previous
section, we derive the linear EL equation for $b(z)$ from Lagrangian (\ref%
{LL}), split it in the real and imaginary parts by substituting $%
b(z)=b_{1}(z)+ib_{2}(z)$, and look for solutions with $b_{1,2}(z)\sim \exp
\left( \lambda z\right) $, cf. Eqs. (\ref{EL})-(\ref{lambda}). Skipping
technical details, the VA predicts the onset of instability of the
antisymmetric state, accounted for by the emergence of $\lambda ^{2}>0$,
when norm $N$ of the stationary antisymmetric solutions exceeds the
respective critical value, $N_{\mathrm{ce}}$. In particular, this value
takes a simple form if one uses the simplified ansatz (\ref{uu}) with $%
A_{3}=0$, when the parameters of the antisymmetric state are $%
A_{1}^{2}=N/\pi $ and $k=-1/2-3N/(4\pi )$:%
\begin{equation}
N_{\mathrm{ce}}=\frac{2\pi ^{2}+4\pi \epsilon }{3\pi +20\epsilon +3\pi
\epsilon ^{2}}.  \label{Nce}
\end{equation}%
This prediction is compared to numerical results below in Fig. \ref{fig9}.

\subsection{Numerical findings}

\begin{figure}[h]
\begin{center}
\includegraphics[height=3.7cm]{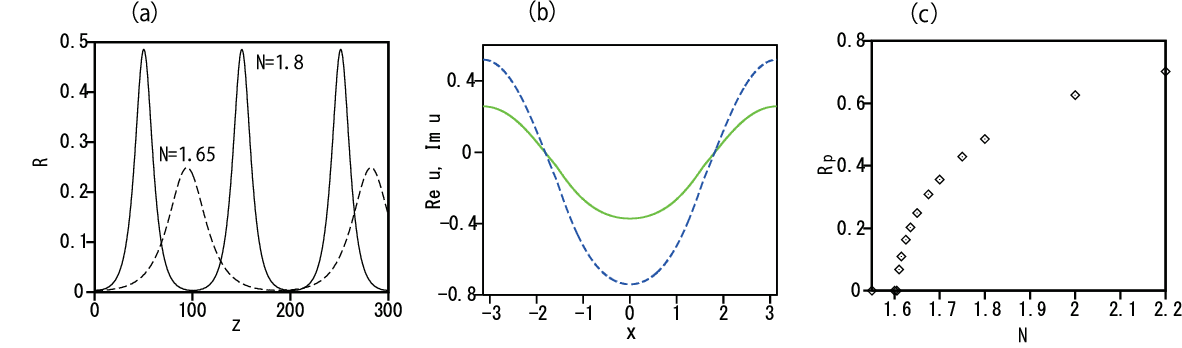}
\end{center}
\caption{(a) The evolution of the asymmetry measure (\protect\ref{R}) of the
unstable antisymmetric (lowest-ES) modes, produced by simulations of Eq. (%
\protect\ref{NLS}) with $\protect\sigma =-1$ and $\protect\epsilon =0.5$,
for two values of the total norm: $N=1.80$ and $1.65$ (solid and dashed
lines, respectively). (b) Re$\left( u(x,t)\right) $ and Im$\left(
u(x,t)\right) $ (the solid green and dashed blue lines, respectively) at $%
z=50$ for $N=1.8$. (c) The maximum (peak) value of the asymmetry measure, $%
R_{\mathrm{p}}$, vs. the total norm, $N$, for the unstable antisymmetric
states at $\protect\epsilon =0.5$.}
\label{fig8}
\end{figure}

An example of the instability of the lowest ES (antisymmetric mode) is
demonstrated in Fig. \ref{fig8}(a) with the help of direct simulations of
Eq. (\ref{NLS}) with $\sigma =-1$ and $\epsilon =0.5$, for two cases, with
total norms $N=1.65$ and $1.80$. Similar to Fig. \ref{fig5}(a), the
oscillatory regime initiated by the instability is plotted by means of the
respective dependences of the asymmetry measure (\ref{R}) on $z$ (note that $%
R=0$ for symmetric and antisymmetric states alike). Figure \ref{fig8}(b)
illustrates the shape of the oscillatory state by means of snapshots of Re$%
\left( u(x,t)\right) $ and Im$\left( u(x,t)\right) $ for $N=1.80$ at $z=50$,
when the asymmetry measure takes a large value, $R=0.4853$. The picture
clearly demonstrates that the antisymmetry is indeed broken, as the zero
points are shifted from $x=\pm \pi /2\approx \pm \allowbreak 1.\,\allowbreak
571$ to $x\approx \pm 1.815$. Note that values of $\left\vert
u(x,z)\right\vert $ at points $x=\pm \pi /2$ are nonzero but small in this
configuration, therefore the corresponding jumps of $\partial u/\partial x$
at these points are inconspicuous in Fig. \ref{fig8}(b).

For the parameters fixed in Figs. \ref{fig8}(a,b) ($\sigma =-1$, $\epsilon
=0.5$), the SASB\ transition is systematically displayed in Fig. \ref{fig8}%
(c) by dint of the dependence of the maximum (peak) value $R_{\mathrm{p}}$,
of the oscillating asymmetry parameter $R$, on $N$. The gradual increase of $%
R_{\mathrm{p}}$ with $N$ demonstrates a continuous SASB transition, cf. Fig. %
\ref{fig3}(b) for the SSB transition in the self-focusing system. On the
other hand, on the contrary to the usual situation with SSB, which occurs
under the action of the self-focusing, Eqs. (\ref{NLS}) and (\ref{U}) do not
produce any stationary state with broken antisymmetry at $N>N_{\mathrm{ce}}$
(above the SASB threshold). Thus, the conclusion is that the SASB transition
leads to the establishment of the robust oscillating states (breathers).

\begin{figure}[h]
\begin{center}
\includegraphics[height=4.2cm]{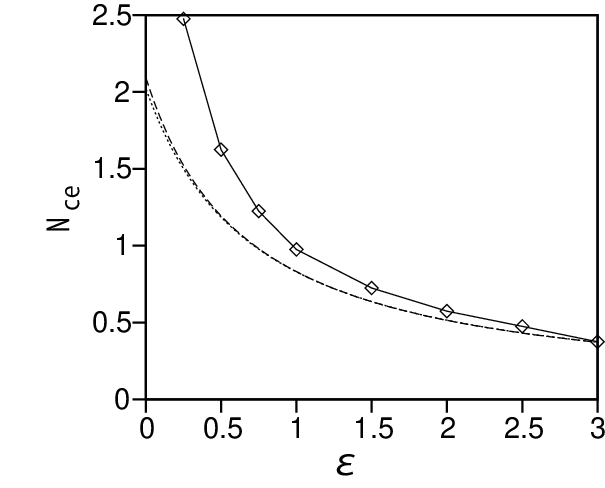}
\end{center}
\caption{Rhombuses represent the critical value $N_{\mathrm{ce}}$ for the
onset of the antisymmetry-breaking instability as a function of $\protect%
\epsilon $, obtained from the numerical solution of Eq. (\protect\ref{NLS})
with $\protect\sigma =-1$. The nearly identical dashed and dotted lines
represent the values of $N_{\mathrm{ce}}$ as predicted by VA with $A_{3}=0$
and $A_{3}\neq 0$, respectively, in ansatz (\protect\ref{uu}).}
\label{fig9}
\end{figure}

Finally, the dependence of the critical norm $N_{\mathrm{ce}}$, at which
SASB commences, on $\epsilon $ is plotted in Fig. \ref{fig9}. The chain of
rhombuses represents numerically found values of $N_{\mathrm{ce}}$, while
the dashed curve corresponds to Eq. (\ref{Nce}), which is predicted by the
simplified VA, with $A_{3}=0$ in ansatz (\ref{uu}) (the result of the full
VA, including $A_{3}\neq 0$, is presented too, being virtually the same). As
well as the dependence $N_{\mathrm{c}}(\epsilon )$ for the SSB, which is
plotted above in Fig. \ref{fig4}, the present one demonstrate the decrease
of the critical norm with the increase of $\epsilon $, which is explained by
the same reason: the weaker coupling between the half-ring boxes, in the
case of larger $\epsilon $, allows weaker defocusing nonlinearity to
initiate SASB. The discrepancy between the numerically found and
VA-predicted values of $N_{\mathrm{ce}}$, observed in Fig. \ref{fig9}, is
explained by insufficient accuracy of ansatz (\ref{uu}).


\section{Conclusion}

The objective of this work is to introduce a basic setting for the
implementation of SSB (spontaneous symmetry breaking) and SASB (spontaneous
antisymmetry breaking) under the action of the self-focusing and defocusing
cubic nonlinearity, respectively, in the ring split in two boxes by
delta-functional potential barriers with strength $\epsilon $, set at
diametrically opposite points. The setting can be implemented in optics and
for matter waves in BEC. First, the spectrum of the linearized model was
found in the approximate analytical and full numerical forms. In the
framework of the nonlinear system, both SSB and SASB were predicted
sufficiently accurately by the VA (variational approximation), and were
studied systematically by means of numerical methods. In addition, a
particular exact analytical solution with strong asymmetry was found, and
the approximation for large $\epsilon $ was presented. In the case of the
self-focusing nonlinearity, the SSB-initiating instability of the spatially
symmetric GS (ground state) is actually the MI (modulational instability)
driven by the self-focusing. The SSB gives rise to the supercritical
bifurcation, which creates stable stationary asymmetric states. In the case
of the defocusing nonlinearity, the symmetric GS remains stable, while SASB
destabilizes the lowest spatially antisymmetric ES (excited state),
replacing it by a robust oscillatory mode with broken antisymmetry.

As an extension of the work, it may be interesting to introduce a similar
two-component model, based on a pair of linearly or nonlinearly coupled NLS
equations, cf. the two-component system in the potential box split by the
central delta-functional barrier \cite{Acus}.

\section*{Acknowledgments}

The work of B.A.M. was supported, in part, by the Israel Science Foundation
through grant No. 1695/22.

\section*{Data availability}

Details of numerical data can be provided on a reasonable request.

\end{document}